\documentclass[fleqn,usenatbib]{mnras}

\usepackage{newtxtext,newtxmath}

\usepackage[T1]{fontenc}

\DeclareRobustCommand{\VAN}[3]{#2}
\let\VANthebibliography\thebibliography
\def\thebibliography{\DeclareRobustCommand{\VAN}[3]{##3}\VANthebibliography}

\usepackage{graphicx}	
\usepackage{amsmath}	
\usepackage{amsfonts}

\def\prd{{\em Phys. Rev. }{\bf D }}

\newcommand{\lya}{Ly$\alpha$}
\newcommand{\lyaf}{Ly$\alpha$ forest}
\newcommand{\lyb}{Ly$\beta$}
\newcommand{\lyalya}{Ly$\alpha \times$Ly$\alpha$}
\newcommand{\lyaqso}{Ly$\alpha \times$QSO}
\newcommand{\qsoqso}{QSO$ \times$QSO}

\newcommand{\lcdm}{{$\Lambda$CDM}}

\usepackage[pagewise,displaymath, mathlines]{lineno}

\title[Beyond BAO with the 3D Lyman-$\alpha$ forest]{Cosmology beyond BAO from the 3D distribution of the Lyman-$\alpha$ forest}

\author[A. Cuceu et al.]{
Andrei Cuceu,$^{1}$\thanks{E-mail: andrei.cuceu.14@ucl.ac.uk}
Andreu Font-Ribera,$^{2,1}$
Benjamin Joachimi$^{1}$
and Seshadri Nadathur$^{1}$
\\
$^{1}$Department of Physics and Astronomy, University College London, Gower Street, London WC1E 6BT, United
Kingdom\\
$^{2}$Institut de F\'{\i}sica d'Altes Energies, The Barcelona Institute of Science and Technology, Campus UAB, 08193 Bellaterra
(Barcelona), Spain\\
}

\date{Accepted XXX. Received YYY; in original form ZZZ}

\pubyear{2015}

\begin{document}
\label{firstpage}
\pagerange{\pageref{firstpage}--\pageref{lastpage}}
\maketitle

\begin{abstract}
We propose a new method for fitting the full-shape of the Lyman-$\alpha$ (\lya) forest three-dimensional (3D) correlation function in order to measure the Alcock-Paczynski (AP) effect. Our method preserves the robustness of baryon acoustic oscillations (BAO) analyses, while also providing extra cosmological information from a broader range of scales. We compute idealized forecasts for the Dark Energy Spectroscopic Instrument (DESI) using the \lya\ auto-correlation and its cross-correlation with quasars, and show how this type of analysis improves cosmological constraints. The DESI \lya\ BAO analysis is expected to measure $H(z_\mathrm{eff})r_\mathrm{d}$ and $D_\mathrm{M}(z_\mathrm{eff})/r_\mathrm{d}$ with a precision of $\sim0.9\%$, where $H$ is the Hubble parameter, $r_\mathrm{d}$ is the comoving BAO scale, $D_\mathrm{M}$ is the comoving angular diameter distance and the effective redshift of the measurement is $z_\mathrm{eff}\simeq2.3$. By fitting the AP parameter from the full shape of the two correlations, we show that we can obtain a precision of $\sim0.5-0.6\%$ on each of $H(z_\mathrm{eff})r_\mathrm{d}$ and $D_\mathrm{M}(z_\mathrm{eff})/r_\mathrm{d}$. Furthermore, we show that a joint full-shape analysis of the \lya\ auto and cross-correlation with quasars can measure the linear growth rate times the amplitude of matter fluctuations in spheres of $8\;h^{-1}$Mpc, $f\sigma_8(z_\mathrm{eff})$. Such an analysis could provide the first ever measurement of $f\sigma_8(z_\mathrm{eff})$ at redshift $z_\mathrm{eff}>2$. By combining this with the quasar auto-correlation in a joint analysis of the three high-redshift two-point correlation functions, we show that DESI could be able to measure $f\sigma_8(z_\mathrm{eff}\simeq2.3)$ with a precision of $5-12\%$, depending on the smallest scale fitted.
\end{abstract}




\section{Introduction}

The vast amount of cosmological data from spectroscopic surveys is usually compressed into summary statistics such as the correlation function or power spectrum. These statistics can be directly used to measure cosmological parameters; however, it is common to split the inference into two steps \citep[e.g.][]{Beutler:2011,Ross:2015,Alam:2017,eBOSS:2020}. A template is first used to model the power spectrum or correlation function in order to measure a few relevant quantities that contain most of the cosmological information. These measurements are then used to fit cosmological parameters for some model, for example flat $\Lambda$ Cold Dark Matter (\lcdm), in combination with other probes, usually the cosmic microwave background \citep[e.g. from][]{PlanckCollaboration:2018}. This approach is used because it contains minimal assumptions, and the full two-point statistic is compressed into a few well understood physical quantities.

Measuring the scale of the acoustic peak from the baryon acoustic oscillation (BAO) signal is one of the most widely used compression methods. This is usually done by splitting the template into a peak and a smooth component for the correlation function, or wiggles and no-wiggles components for the power spectrum. The coordinates of the peak (or wiggles) component are then re-scaled in order to fit the BAO scale from the data. This method has been used to measure the BAO scale using the galaxy distribution at redshifts $z\lesssim1$ \citep[e.g.][]{Eisenstein:2005,Cole:2005}, the quasar (QSO) distribution at redshifts $0.8<z<2.2$ \citep[e.g.][]{Ata:2018}, and the Lyman-$\alpha$ (\lya) forest at redshifts $2<z<3$ \citep[e.g.][]{Busca:2013,Slosar:2013,Kirkby:2013,FontRibera:2014}.

The \lyaf\ consists of a series of absorption lines blueward of the \lya\ emission peak in spectra of high-redshift quasars \citep[e.g.][]{Lynds:1971,Rauch:1998}. The forest appears due to absorption by neutral hydrogen between the quasar and us, which means it traces the intergalactic medium. This makes it a great tool for cosmology as it probes the distribution of matter at redshifts ($z\gtrsim2$) that are generally hard to access with other probes \citep[see e.g.][for early cosmological applications]{Croft:1999,mcdonald:2000,Croft:2002,Viel:2004}.

A common way to extract more information from the two-point statistics of discrete tracers is to fit the full shape (instead of just the peak component) in order to measure the growth rate of structure through redshift space distortions \citep[RSD; e.g][]{Blake:2011,Reid:2012,Beutler:2012,Samushia:2014}. This approach is not possible for the \lyaf\ because we have to marginalize over an unknown velocity gradient bias which is degenerate with the growth rate. This bias appears because we work with the two-point statistics of flux, which has a non-linear mapping to the directly distorted field of optical depth \citep[see e.g.][]{Slosar:2011,McDonald:2003,Givans:2020,Chen:2021}. Therefore, an RSD analysis using the \lyaf\ three dimensional (3D) correlation function has so far been out of reach.

The analysis of the \lya\ 3D auto-correlation function (\lyalya) and its cross-correlation with the quasar distribution (\lyaqso) has evolved considerably since they were first used to measure the BAO peak from Baryon Oscillation Spectroscopic Survey (BOSS) data \citep{Busca:2013,Slosar:2013,Kirkby:2013,FontRibera:2014}. A physical model for the correlations was introduced by \cite{Bautista:2017} and \cite{duMasdesBourboux:2017}. This model includes the effect of metal line contamination and that of high column density (HCD) systems. With the extended BOSS (eBOSS) analyses, the \lya\ signal from the \lyb\ section of the forest was also used, first through its correlation with \lya\ signal in the \lya\ section \citep{DeSainteAgathe:2019}, and then through its correlation with the QSO distribution \citep{Blomqvist:2019,Bourboux:2020}. Even though major advancements have been made in modeling and understanding the 3D \lyalya\ and \lyaqso\ statistics, so far they have only been used to measure BAO.

In this work we investigate the potential for extracting more cosmological information from the 3D distribution of the \lyaf\ through the Alcock-Paczynski (AP) effect \citep{Alcock:1979,Hui:1999,McDonald:1999,McDonald:2003}. This appears due to the choice of fiducial cosmology which is used to transform the measured angles and redshifts into comoving coordinates. If this fiducial cosmology is different from the true cosmology, the measured correlation will have an extra anisotropy. Thus isolating this anisotropic AP contribution allows us to determine the true background cosmological model. Some of this AP signal is measured through anisotropic BAO analyses, by measuring two distinct scales along versus across the line of sight. However, this distortion affects the whole correlation function. Therefore, the first objective of this article is to complement standard \lya\ BAO analyses with AP constraints from a broader range of scales.


The two \lyaf\ correlation functions (\lyalya\ and \lyaqso) are some of our best probes of the Universe at redshifts $1.8<z<4$. However, there is big potential for a third correlation function in this redshift range: the quasar auto-correlation (\qsoqso). As mentioned above, this has already been used to measure both BAO and the growth rate of structure at effective redshifts $z_\mathrm{eff}\simeq1.6$. With the start of the Dark Energy Spectroscopic Instrument (DESI) survey, we will have new quasar catalogues with about 0.7 million expected to be at redshifts $z>2.1$ \citep{Aghamousa:2016}. This opens up the potential of performing a joint analysis of the three correlation functions (\lyalya, \lyaqso\ and \qsoqso) for the first time. Jointly fitting the full shape of all three correlations would allow us to take full advantage of the synergies between them, and lead to more precise and robust constraints. Our second goal in this work is to investigate how such an analysis could be performed and study its benefits, including the potential for measuring RSD.

We start by introducing our methodology for template-fitting the full shape of the \lyaf\ correlation function in Section \ref{sec:theory}. We also compare our approach with that used in past analyses of discrete tracers. After that, in Section \ref{sec:lya_ap} we perform a forecast analysis to demonstrate how the AP effect can be measured from the full shape of the correlation while preserving the robust BAO measurement. We also demonstrate the usefulness of such a measurement in constraining cosmological parameters in a flat \lcdm\ model. Finally, in Section \ref{sec:3x2pt} we forecast a joint analysis of the three high-redshift two-point (high$-z$ $3\times2$pt) correlation functions (\lyalya, \lyaqso\ and \qsoqso) in order to study their synergies and showcase the potential benefits of such an analysis.

\section{Method}
\label{sec:theory}

Our model of the 3D correlation function is based on the framework introduced by \cite{Kirkby:2013} and used in all \lyaf\ BAO analyses. Our approach is meant to extend these analyses to also include information from the broadband. We use a template power spectrum and introduce parameters that re-scale its coordinates. A fit to the data allows us to place constraints on these scale parameters. The resulting measurements can be transformed into constraints on cosmological parameters. We start by introducing these scale parameters in Section \ref{subsec:scale}. After that, we introduce the components of the template in Section \ref{subsec:two_comp}, and compare our approach to BAO analyses and previous full-shape analyses. In Section \ref{subsec:cfmodel}, we introduce our models for the \lyaf\ auto-correlation, its cross-correlation with quasars, and the quasar auto-correlation. Finally, in Section \ref{subsec:comparison} we showcase the effects of our scale parameters on the \lyaf\ correlation function.

\subsection{Scale parameters}
\label{subsec:scale}

When computing the 3D correlation function, we transform the observed redshift and angular separations $(\Delta z,\Delta\theta)$ into comoving coordinates $(r_{||},r_{\bot})$. For positions $i$ and $j$, at redshifts $z_i$ and $z_j$ and separated by an angle $\Delta\theta$, we define the radial coordinates as \citep{Bourboux:2020}
\begin{linenomath*}
\begin{equation}
\begin{aligned}
    r_{||} = [D_\mathrm{C,fid}(z_i) - D_\mathrm{C,fid}(z_j)] \cos{\frac{\Delta\theta}{2}}, \\
    r_{\bot} = [D_\mathrm{M,fid}(z_i) + D_\mathrm{M,fid}(z_j)] \sin{\frac{\Delta\theta}{2}},
\end{aligned}
\end{equation}
\end{linenomath*}
where $D_\mathrm{M}(z)$ is the comoving angular diameter distance and $D_\mathrm{C}(z)=c\int_0^z \mathrm{d}z/H(z)$ is the radial comoving distance, with $c$ as the speed of light and $H(z)$ as the Hubble parameter. The \textit{fid} term indicates that we use a fiducial cosmology to compute these distances. If the true cosmology is different from the fiducial one, the ratio between the inferred line of sight and transverse distances will be different from the true ratio. This means we will observe an apparent anisotropy in the measured correlation, which is the Alcock-Paczynski effect we wish to measure \citep{Alcock:1979}. Note however, that there are other sources of anisotropy, such as RSD. In order to measure the AP effect, we have to correctly model and marginalize over all other anisotropies.

When building a model for the correlation function, we follow past \lyaf\ BAO analyses and use a template power spectrum computed using a fixed cosmology. Following \cite{Kirkby:2013}, we allow for small differences between the template and measured cosmologies by using general coordinate transformations of the form $r_{||} \xrightarrow{} r_{||}'(r_{||},r_{\bot},z)$ and $r_{\bot} \xrightarrow{} r_{\bot}'(r_{||},r_\bot,z)$.

The most commonly used parametrisation for anisotropic re-scalings is given by:
\begin{linenomath*}
\begin{equation}
    r_{||}'=q_{||}r_{||}, \;\;\;\; r_\bot'=q_\bot r_\bot,
\end{equation}
\end{linenomath*}
where $(q_{||},q_\bot)$ re-scale the coordinates along and across the line of sight respectively. However, we wish to isolate the AP effect which changes the ratio $r_\bot/r_{||}$. Therefore, we define the parameters:
\begin{linenomath*}
\begin{equation}
    \phi(z) \equiv \frac{q_\bot(z)}{q_{||}(z)} \;\;\text{and}\;\; \alpha(z) \equiv \sqrt{q_\bot(z) q_{||}(z)},
\end{equation}
\end{linenomath*}
where $\phi(z)$ re-scales the ratio: $r_\bot'/r_{||}'=\phi \; r_\bot/r_{||}$, and is meant to measure the AP effect. On the other hand, $\alpha(z)$ re-scales the product $r_\bot' r_{||}'=\alpha^2 \; r_\bot r_{||}$, which translates into an isotropic re-scaling of $\xi$. The effect of these parameters becomes clearer when we consider their impact on the radial and transverse coordinates through small deviations around $\phi=1,\alpha=1$:
\begin{linenomath*}
\begin{equation}
\begin{aligned}
    r_{||}' = \alpha r_{||} - \frac{\phi - 1}{2} r_{||} + \mathcal{O}[(\phi - 1)^2, (\alpha - 1)^2], \\
    r_\bot' = \alpha r_\bot + \frac{\phi - 1}{2} r_\bot + \mathcal{O}[(\phi - 1)^2, (\alpha - 1)^2].
\end{aligned}
\end{equation}
\end{linenomath*}
The $\alpha$ parameter produces the same effect on both $r_{||}$ and $r_\bot$, which corresponds to isotropic re-scaling. On the other hand, $\phi$ produces small changes that are directly opposite in $r_{||}$ versus $r_\bot$, which corresponds to anisotropy in $\xi$. We study the effect of these parameters on the correlation function in \ref{subsec:comparison}.

These quantities are an intermediate step between fitting the correlation function and constraining cosmological parameters. Having defined the scale parameters we will use, we turn our attention to the template and the application of these parameters.


\subsection{Two-component full-shape parametrisation}
\label{subsec:two_comp}

We construct our model based on the separation of the BAO feature from the rest of the correlation. This is achieved by starting with a template isotropic linear power spectrum for an assumed fiducial cosmology, computed using CAMB \citep{Lewis:1999}. This template power spectrum is decomposed into a peak (or wiggles) component, $P_\text{peak}(k,z_\text{eff})$, and a smooth (or no-wiggles) component, $P_\text{smooth}(k,z_\text{eff})$, using the method described in \cite{Kirkby:2013}. The reason for this separation is that BAO is a clear feature that can be used as a standard ruler; it has been studied extensively and we know that for the \lyaf\ it is very robust when it comes to contaminants \citep[e.g.][]{Cuceu:2020}. Therefore we consider it advantageous to separate this feature, because it will make it easier to study and understand the information contained in the rest of the correlation (i.e. in the broadband), and how it is affected by contaminants (e.g. HCDs and continuum fitting).

The full transformed correlation in the original coordinates ($r_{||},r_\bot,z$) is given by:
\begin{linenomath*}
\begin{equation}
    \xi_\text{full}(r_{||},r_\bot,z) = \xi_\text{peak}(r_{||}',r_\bot',z) + \xi_\text{smooth}(r_{||}'',r_\bot'',z),
\end{equation}
\end{linenomath*}
where the transformed coordinates of the peak component $(r_{||}',r_\bot')$ are allowed to be different from the transformed coordinates of the smooth component $(r_{||}'',r_\bot'')$. For comparison, in BAO analyses we would fix the smooth component: $(r_{||}'',r_\bot'') = (r_{||},r_\bot)$, whereas past full-shape analyses did not use the peak-smooth decomposition, which would be equivalent to fixing the two sets of transformations to be the same: $(r_{||}'',r_\bot'') = (r_{||}',r_\bot')$.

As we have two sets of coordinate transformations, we will need two sets of $(\phi,\alpha)$ parameters. The AP effect distorts the entire correlation, and $\phi$ is meant to measure this anisotropy. Therefore, both the smooth and peak components are affected by $\phi$ in the same way. This means that we would ideally sample only one $\phi$ parameter that re-scales both components. However, as we wish to understand the cosmological value added by re-scaling the broadband, and also study how each parameter is affected by contaminants, we will keep them separate. Going forward we will use the notation $\phi_\mathrm{s}$ for the smooth component and $\phi_\mathrm{p}$ for the BAO peak component. A measurement of $\phi$ corresponds to a measurement of:
\begin{linenomath*}
\begin{equation}
	\textbf{AP: } \; \phi(z) = \frac{F_\text{AP}(z)}{F^\text{fid}_\text{AP}(z)} = \frac{D_\mathrm{M}(z) H(z)}{[D_\mathrm{M}(z) H(z)]_\text{fid}},
	\label{eq:phi}
\end{equation}
\end{linenomath*}
where the AP parameter is defined as the ratio of two distances $F_\mathrm{AP}(z)=D_\mathrm{M}(z)/D_\mathrm{H}(z)$, with $D_\mathrm{H}(z)=c/H(z)$.

On the other hand, the $\alpha$ parameter has different interpretations for the peak and smooth components. We not only need to account for the different expansion histories between the template and the data, but also for the features that set the scale we measure. We denote the parameter that isotropically re-scales the peak as $\alpha_\mathrm{p}$ and the equivalent parameter for the smooth component as $\alpha_\mathrm{s}$. In the case of the BAO peak, the relevant scale is the size of the sound horizon at the end of the drag epoch, $r_\mathrm{d}$. The isotropic scale of the peak component, $\alpha_\mathrm{p}$, corresponds to a measurement of:
\begin{linenomath*}
\begin{equation}
    \textbf{BAO: } \; \alpha_\mathrm{p}(z) = \sqrt{\frac{D_\mathrm{M}(z) D_\mathrm{H}(z) / r_\mathrm{d}^2}{[D_\mathrm{M}(z) D_\mathrm{H}(z) / r_\mathrm{d}^2]_\text{fid}}}.
    \label{eq:alpha}
\end{equation}
\end{linenomath*}

On the other hand, $\alpha_\mathrm{s}$ is harder to identify with one clear feature. The scale of matter-radiation equality ($k_\mathrm{eq}$) is a feature that contributes to $\alpha_\mathrm{s}$, and has successfully been used to constrain cosmology from the power spectrum \citep{Baxter:2021,Philcox:2020}. However, it is not clear that it is the only feature that contributes to the isotropic scale of the broadband. Furthermore, the effect produced by $\alpha_\mathrm{s}$ is very similar to that of the \lya\ flux bias, which could lead to the two parameters being hard to disentangle. Therefore, we will not focus on the cosmological interpretation of $\alpha_\mathrm{s}$ in this work, and leave it to future studies to determine if this parameter could be useful.

In past galaxy full-shape analyses there was no smooth/peak decomposition, and the isotropic scale parameter was interpreted using $r_\mathrm{d}$ \citep[e.g.][]{Beutler:2017}. This is based on the approximation that most of the signal comes from the BAO peak. This means we can measure $\alpha_\mathrm{p}$ very precisely, but not $\alpha_\mathrm{s}$, so the measurement of a parameter $\alpha=\alpha_\mathrm{p}=\alpha_\mathrm{s}$ would be dominated by signal from the peak. By fitting two different parameters we will be able to test this assumption.

Finally, for clarity we show how our new set of parameters $(\phi_\mathrm{s}, \alpha_\mathrm{s}, \phi_\mathrm{p}, \alpha_\mathrm{p})$ would be treated in BAO and galaxy full-shape analyses:
\begin{linenomath*}
\begin{equation}
\begin{aligned}
    \text{Standard BAO analyses: } & (\phi_\mathrm{s}, \alpha_\mathrm{s}) \text{ fixed to } (1, 1), \\
    \text{Galaxy full-shape: } & (\phi_\mathrm{s}, \alpha_\mathrm{s}) \text{ fixed to }  (\phi_\mathrm{p}, \alpha_\mathrm{p}), \\
    \text{Two-component \lya} \;\;\;\;\; &  \\
    \text{full-shape (this work): } & (\phi_\mathrm{s}, \alpha_\mathrm{s}, \phi_\mathrm{p}, \alpha_\mathrm{p}) \text{ all free}.
\end{aligned}
\end{equation}
\end{linenomath*}

In the rest of this work, we show the effects of $\phi_\mathrm{s}$ and $\alpha_\mathrm{s}$ on the correlation function, study the potential for measuring them using the \lyaf\ and its cross-correlation with quasars, and show their usefulness for constraining cosmology. However, we leave it to future work to investigate how they interact with contaminants and potential systematic errors that may affect them.

\subsection{Correlation function model}
\label{subsec:cfmodel}

\begin{figure*}
    \includegraphics[width=1\textwidth,keepaspectratio]{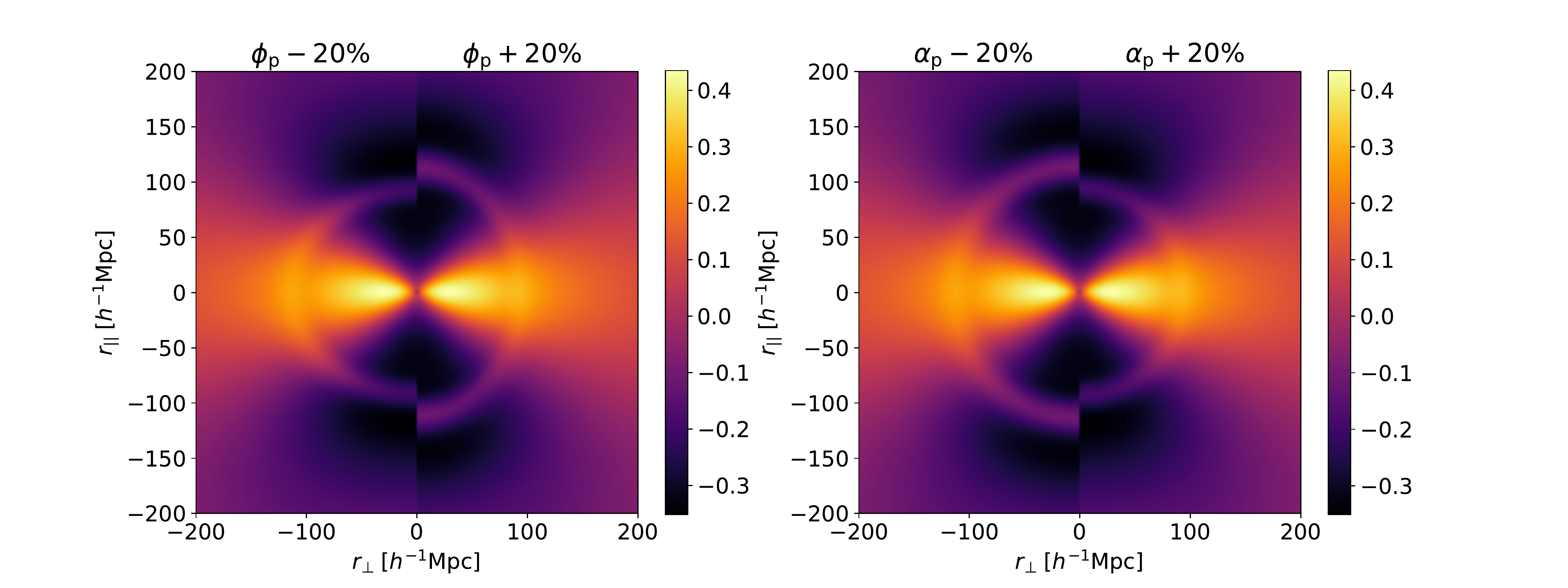}
    \includegraphics[width=1\textwidth,keepaspectratio]{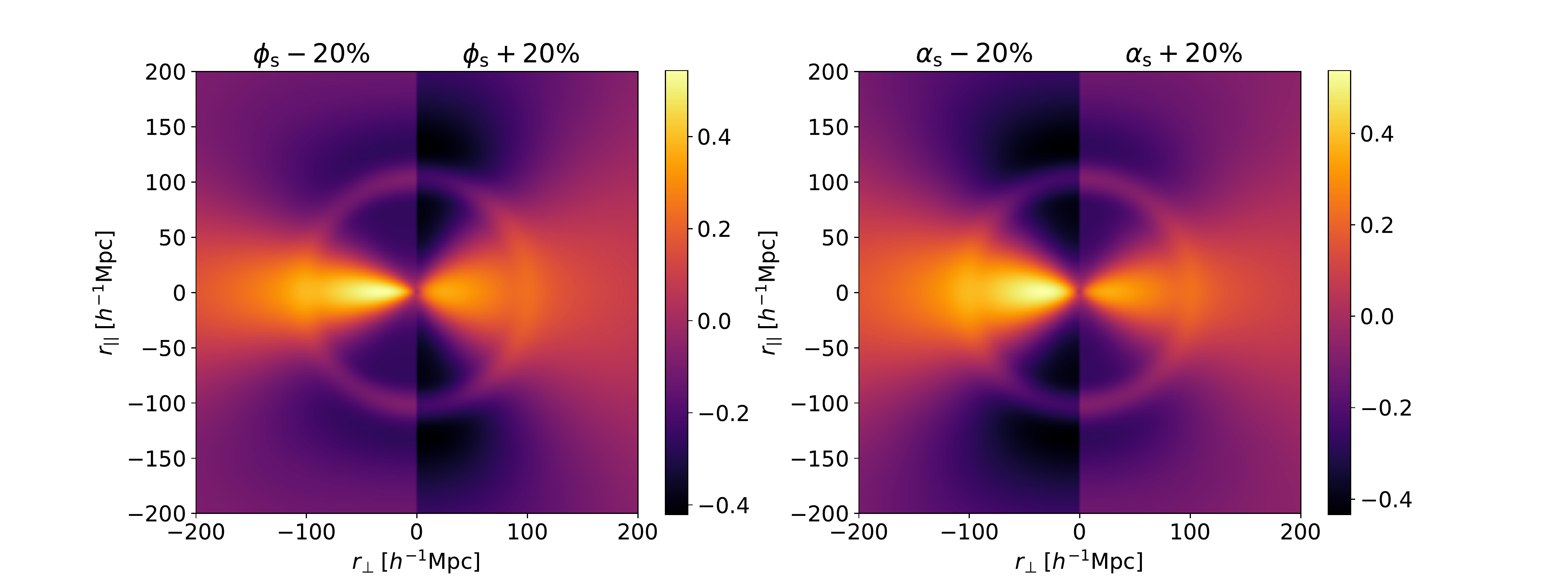}
    \includegraphics[width=1\textwidth,keepaspectratio]{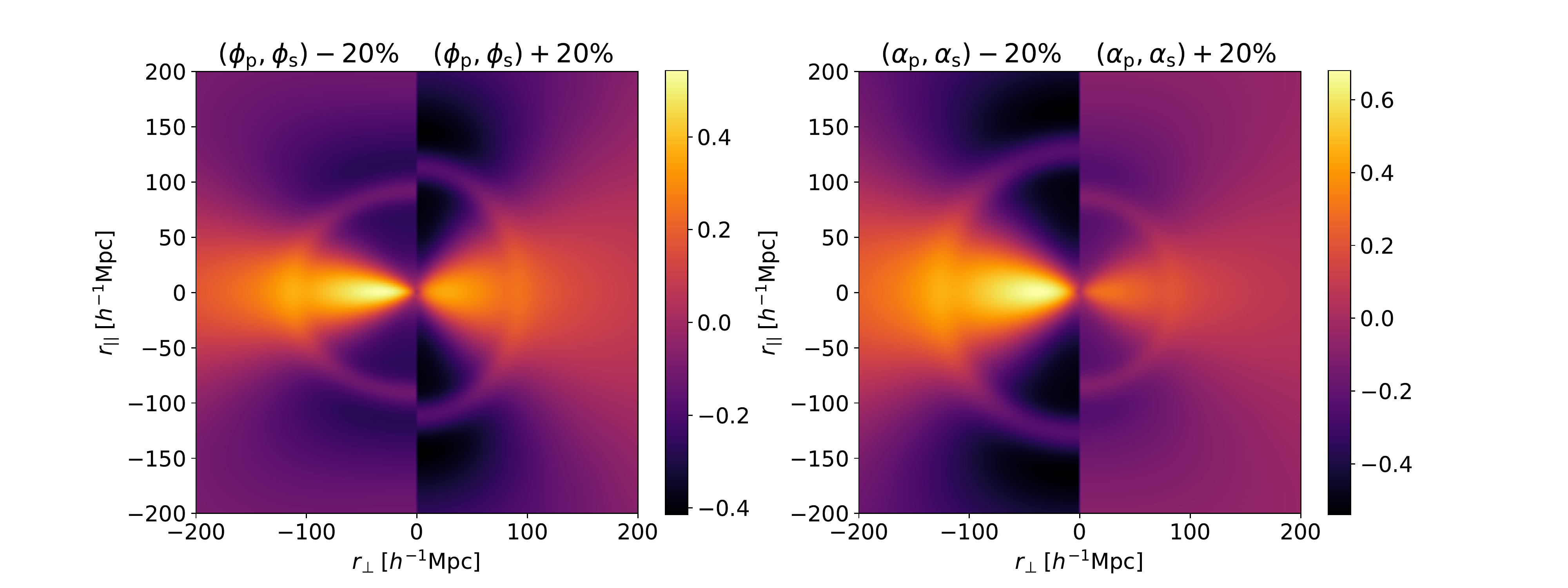}
    \caption{Contour plots of the \lyaf\ auto-correlation function $(r^2\xi)$ in terms of the radial coordinates along and across the line of sight $(r_{||},r_\bot)$. Each plot shows the correlation function computed using a smaller value of the given parameter on the left and a higher value on the right. The left column shows the effect of the $\phi$ parameters which change the anisotropy of the correlation (the AP effect). The right column shows the effect of the $\alpha$ parameters which change the isotropic scale of the correlation. The top row shows the scale parameters for the peak component ($\phi_\mathrm{p}$ and $\alpha_\mathrm{p}$). This is what BAO analyses measure. The middle row shows the parameters that re-scale the smooth component ($\phi_\mathrm{s}$ and $\alpha_\mathrm{s}$). We aim to measure both the BAO parameters and the broadband parameters. Finally, the bottom row shows the effect of changing the parameters for the peak and smooth components at the same time ($\phi_\mathrm{s}=\phi_\mathrm{p}$ and $\alpha_\mathrm{s}=\alpha_\mathrm{p}$). This is what past spectroscopic galaxy clustering analyses (e.g. BOSS and eBOSS) measure when fitting the full shape of the correlation.}
    \label{fig:cf2d}
\end{figure*}



Our models for the \lyaf\ auto-correlation and its cross-correlation with quasars follow \cite{Bourboux:2020}, however we use simplified versions with no contaminants or distortion due to the effect of continuum fitting. The \lyaf\ analyses of the auto and cross-correlation have so far only been done using models with linear-order perturbations. For \lyalya, a small scale non-linear correction term is also used, with the parameter values calibrated using simulations \citep{Arinyo:2015}. On the other hand, full-shape analyses of \qsoqso\ typically use higher-order perturbation theory \citep[e.g.][]{Taruya:2010}. In this work we restrict ourselves to linear-order perturbation theory. Therefore, the full anisotropic power spectra of \lyalya\ ($P_{\mathrm{Ly}\alpha}$), \lyaqso\ ($P_{\times}$) and \qsoqso\ ($P_{\mathrm{QSO}}$) are given by:
\begin{linenomath*}
\begin{align}
    P_{\mathrm{Ly}\alpha}(k,\mu_\mathrm{k},z) =\; & b^2_{\mathrm{Ly}\alpha}(1 + \beta_{\mathrm{Ly}\alpha} \mu_\mathrm{k}^2)^2 \; F^2_{\text{nl,Ly}\alpha} P_{\text{fid}}(k,z), 
    \label{eq:kaiser_lya} \\
    P_{\times}(k,\mu_\mathrm{k},z) =\; & b_{\mathrm{Ly}\alpha}(1 + \beta_{\mathrm{Ly}\alpha} \mu_\mathrm{k}^2) \nonumber \times\\ 
    & \times (b_{\mathrm{QSO}} + f(z) \mu_\mathrm{k}^2) \;\; F_\text{nl,QSO} \; P_{\text{fid}}(k,z),
    \label{eq:kaiser_cross} \\
    P_{\mathrm{QSO}}(k,\mu_\mathrm{k},z) =\; & (b_{\mathrm{QSO}} + f(z) \mu_\mathrm{k}^2)^2 \; F^2_\text{nl,QSO} \; P_{\text{fid}}(k,z),
    \label{eq:kaiser_qso}
\end{align}
\end{linenomath*}
where $b_{\mathrm{Ly}\alpha}$ and $b_{\mathrm{QSO}}$ are the linear biases of the \lyaf\ and quasars respectively, $f(z)$ is the logarithmic growth rate, and $\mu_\mathrm{k}=k_{||}/k$, with the wavenumber $k$, and its projection along the line of sight, $k_{||}$.

The RSD parameter of the \lyaf\ is given by:
\begin{linenomath*}
\begin{equation}
    \beta_{\mathrm{Ly}\alpha} = \frac{b_{\eta,\mathrm{Ly}\alpha}f(z)}{b_{\mathrm{Ly}\alpha}},
\end{equation}
\end{linenomath*}
where $b_{\eta,\mathrm{Ly}\alpha}$ is the velocity divergence bias. As $b_{\eta,\mathrm{Ly}\alpha}$ and $f(z)$ always appear together in the \lyaf\ RSD term, they are completely degenerate. Therefore, we use the parameter $\beta_{\mathrm{Ly}\alpha}$ to define the RSD term of the \lyaf, and we marginalize over it instead of $b_{\eta,\mathrm{Ly}\alpha}$. This is meant to separate the degeneracies of different parameter combinations, and to clearly differentiate between nuisance parameters ($b_{\mathrm{Ly}\alpha},b_{\mathrm{QSO}},\beta_{\mathrm{Ly}\alpha}$) and the parameters of interest ($\phi,f$). We also note that the symmetries of Equation \ref{eq:kaiser_cross} mean that when fitting only the cross-correlation, $b_{\mathrm{Ly}\alpha}$ is fully degenerate with $b_{\mathrm{QSO}}$, and similarly $\beta_{\mathrm{Ly}\alpha}$ with $f(z)/b_{\mathrm{QSO}}$.

The small-scale non-linear correction for \lya, $F_{\text{nl,Ly}\alpha}^2$, is given by the model introduced by \cite{Arinyo:2015}. However, this has only been tested and applied to the \lyaf\ auto-correlation, and not for the cross-correlation. Therefore, we only apply this term for \lyalya. On the other hand, the term $F_\text{nl,QSO}$, which models the quasar non-linear velocities, is used for both the cross-correlation and the quasar auto-correlation. Following \cite{Percival:2009}, this is given by:
\begin{linenomath*}
\begin{equation}
    F_\text{nl,QSO}(k_{||}) = \sqrt{\frac{1}{1 + (k_{||} \sigma_\mathrm{v})^2}},
\end{equation}
\end{linenomath*}
where $\sigma_\mathrm{v}$ is a free parameter representing the rms velocity dispersion.

We also model the non-linear broadening of the BAO peak by applying the term $P_\text{nl,peak}$ to the peak component of the power spectrum, $P_\text{peak}(k,z_\text{eff})$, following \cite{Eisenstein:2007}. This term is given by:
\begin{linenomath*}
\begin{equation}
    P_\text{nl,peak} = \exp{[-k_{||}^2\Sigma_{||}^2/2 - k_\bot^2\Sigma_\bot^2/2]},
\end{equation}
\end{linenomath*}
where $k_\bot$ is the projection of the wavenumber $k$ across the line of sight, and the smoothing scales $(\Sigma_{||},\Sigma_\bot)$ are fixed to the values $(6.42,3.26) h^{-1}$Mpc \citep{Kirkby:2013}.

We use the \texttt{Vega} library\footnote{https://github.com/andreicuceu/vega} to compute model correlation functions using the same template power spectrum (and fiducial cosmology) as in \cite{Bourboux:2020}. \texttt{Vega} is a new, improved version of the BAO fitter in the \texttt{picca}\footnote{https://github.com/igmhub/picca} library that was used in eBOSS \lya\ BAO analyses.


\subsection{Impact on the correlation function}
\label{subsec:comparison}

We investigate how the parameters we introduced $(\phi_\mathrm{s}, \alpha_\mathrm{s}, \phi_\mathrm{p}, \alpha_\mathrm{p})$ change the \lyaf\ auto-correlation function using the model presented above. We show the effect produced by these parameters in Figure \ref{fig:cf2d}, using contour plots of the correlation function. For each plot we show a model correlation computed using a $20\%$ smaller value of a given parameter on the left and $20\%$ higher value on the right, while the other parameters are kept fixed to one (scale parameters) or their best fit (nuisance parameters) from \cite{Bourboux:2020}. Note that such changes are extreme, and chosen only to clearly showcase the effect of varying the parameters. The coordinate re-scalings we use are only approximations that work for values close to the template cosmology (parameter values around 1).

The first two rows of Figure \ref{fig:cf2d} show the effect of the four parameters we introduced, with the $\phi$ parameters on the left and the isotropic scale parameters on the right. The top row shows the parameters that only affect the BAO peak, $\phi_\mathrm{p}$ and $\alpha_\mathrm{p}$, while leaving the broadband component mostly unchanged. The former produces an anisotropy in the BAO scale (top left), leading to a different position of the peak along versus across the line of sight. The latter isotropically re-scales the BAO peak (top right). On the other hand, the two plots in the middle row show the parameters that only affect the smooth component, $\phi_\mathrm{s}$ and $\alpha_\mathrm{s}$, while leaving the BAO peak unchanged. $\phi_\mathrm{s}$ changes the anisotropy of the smooth component (middle left). Note that $\xi$ is anisotropic even for $\phi_\mathrm{s}=1$ due to RSD. We will need to marginalize over this effect if we want to measure $\phi_\mathrm{s}$. The $\alpha_\mathrm{s}$ parameter isotropically re-scales the smooth component (middle right) without affecting the position of the BAO peak.

Finally, the bottom row of Figure \ref{fig:cf2d} shows the effect of re-scaling the smooth and peak components at the same time by fixing $\phi_\mathrm{s}=\phi_\mathrm{p}$ (bottom left) and $\alpha_\mathrm{s}=\alpha_\mathrm{p}$ (bottom right). This means that the peak and broadband are entangled, leaving a measurement of $\alpha$ harder to interpret. This is what past full-shape analyses of discrete tracers have measured, but using different parametrisations.

\section{AP Forecasts for the Ly$\alpha$ forest}
\label{sec:lya_ap}

We start our investigation of a potential full-shape analysis from the \lyaf\ by analysing simulated correlation functions. We use these mock correlations to test our proposed two-component full-shape analysis, and forecast how well DESI will be able to measure the four scale parameters we introduced.

\subsection{Mock data}
\label{subsec:data}

\begin{figure*}
    \includegraphics[width=1\textwidth,keepaspectratio]{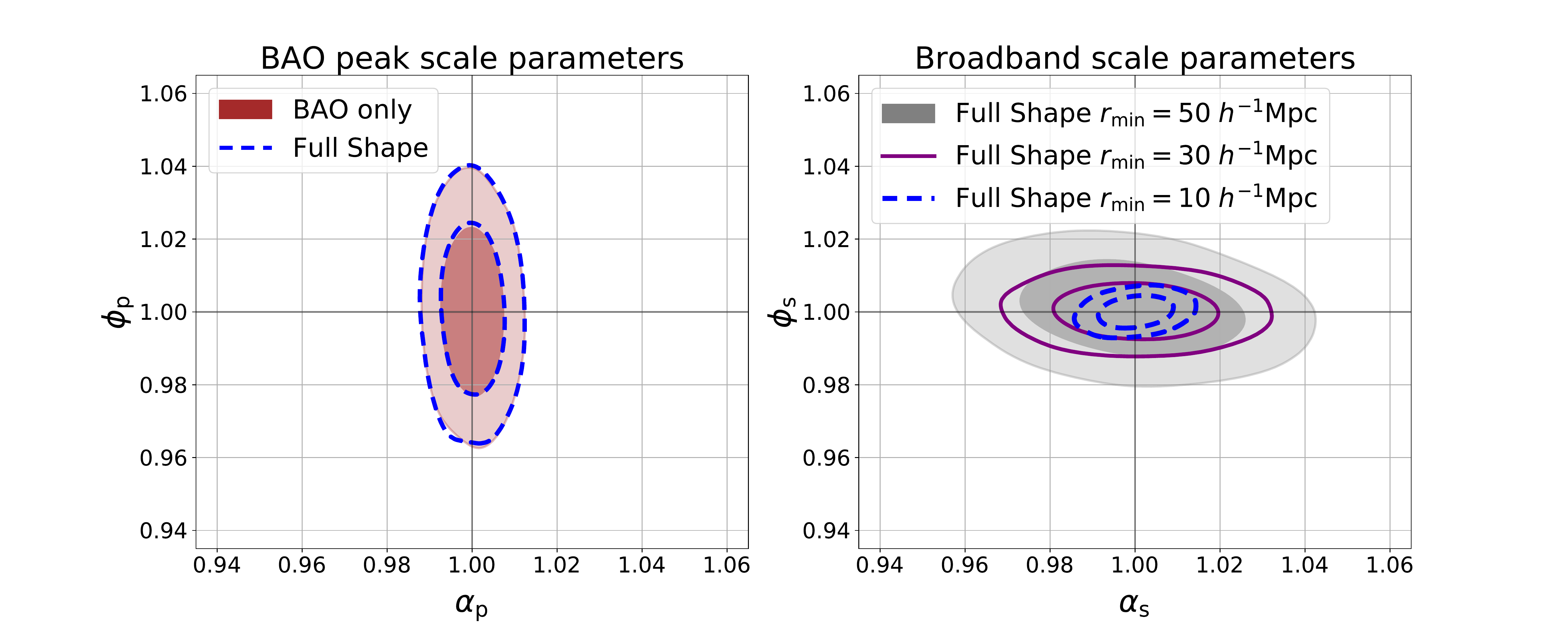}
    \caption{Forecast constraints on the four scale parameters in our two-component full-shape analysis from \lyalya\ and \lyaqso. In the left panel we compare the posterior distributions of the BAO peak parameters $\alpha_\mathrm{p}$ and $\phi_\mathrm{p}$ obtained using a BAO only analysis and our full-shape method. The very good agreement between the two shows that we can isolate the robust BAO information when performing a full-shape analysis. In the right panel we show posterior distributions of the broadband scale parameters $\phi_\mathrm{s}$ and $\alpha_\mathrm{s}$ for different minimum separations used for the fits. This shows that we can obtain much better constraints on the Alcock-Paczynski parameter ($\phi$) from the smooth component compared to those from the BAO peak.}
    \label{fig:scale_params}
\end{figure*}

We compute model correlation functions for \lyalya\ and \lyaqso\ as described in Section \ref{subsec:cfmodel}. The models are computed using the best fit parameter values from eBOSS DR16 ($b_{\mathrm{Ly}\alpha}$=$-0.117$, $\beta_{\mathrm{Ly}\alpha}$=$1.669$, $b_{\mathrm{QSO}}$=$3.73$, $f$=$0.97$, $\sigma_\mathrm{v}$=$6.86 h^{-1}$Mpc), except for the scale parameters, which are all set to equal one. We use these models as our simulated data. For the purposes of this work, we wish to perform a forecast analysis, and therefore we do not add noise to the fiducial data vector.

We use covariance matrices computed from mock data sets by \cite{Farr:2020}. These mocks where created using the \texttt{LyaCoLoRe} package,\footnote{https://github.com/igmhub/LyaCoLoRe} which uses an initial Gaussian random field to simulate \lyaf\ transmitted flux skewers and adds the relevant small scale power and RSD. The mocks were used to create full sky quasar catalogs containing $\sim3.7$ million QSOs above redshift $z=1.8$. These simulated data products were used to compute the \lyaf\ auto-correlation function, its cross-correlations with QSOs, the QSO auto-correlation, and the relevant covariance matrices. In order to compute a covariance matrix relevant for DESI, we use the expected DESI survey area of $14000$ square degrees, and assume it will measure roughly $\sim1.1$ million QSOs above redshift $z=1.8$ \citep{Aghamousa:2016}. We then compute a factor that re-scales the covariance matrix ($f_\text{cov}$) to match the expected DESI number density ($n_\text{DESI}$) and area ($A_\text{DESI}$):
\begin{linenomath*}
\begin{equation}
    f_\text{cov} = \bigg(\frac{n_\text{Mock}}{n_\text{DESI}}\bigg)^2 \frac{A_\text{Mock}}{A_\text{DESI}},
\end{equation}
\end{linenomath*}
where $n_\text{Mock}$ and $A_\text{Mock}$ are the number density and area of the mock correlation computed by \cite{Farr:2020}, and the factor we compute is $f_\text{cov}\simeq4$. This factor is based on the fact that \lyaf\ measurements are still limited by shot noise, and therefore the number density needs to be accounted for alongside the area, which accounts for cosmic variance. We also validate it by comparing our cosmological constraints with the forecasts from \cite{Aghamousa:2016} in Section \ref{subsec:ap_cosmo}. The DESI simulated covariance matrix is then given by $C_\text{DESI} = f_\text{cov} C_\text{Mock}$, based on the mock covariance, $C_\text{Mock}$.

We assume that there is no cross-covariance between \lyalya\ and \lyaqso\ \citep{duMasdesBourboux:2017}, as has been standard with \lya\ BAO analyses so far. We use a Gaussian likelihood, and compute posterior distributions using the Nested Sampler \texttt{PolyChord}\footnote{https://github.com/PolyChord/PolyChordLite} \citep{Handley:2015a,Handley:2015b}. We use the recommended setup (live\textunderscore points = $25\times$ number of parameters, num\textunderscore repeats = $3\times$ number of parameters) when running \texttt{PolyChord}. When fitting each correlation independently, we sample the parameters: $\{\phi_\mathrm{p},\alpha_\mathrm{p},\phi_\mathrm{s},\alpha_\mathrm{s},b_{\mathrm{Ly}\alpha},\beta_{\mathrm{Ly}\alpha}\}$ for the auto-correlation, while the cross-correlation has one extra parameter ($\sigma_\mathrm{v}$). For the cross we do not sample the QSO bias and RSD parameters due to the degeneracies with the \lya\ parameters (see Section \ref{subsec:cfmodel}). When performing joint fits, we also sample $b_\mathrm{QSO}$ and $f(z)$; however, we treat $f(z)$ as a nuisance parameter in this section, and only focus on measuring $\phi_s$.

\subsection{Scale parameters from a two-component full-shape analysis}
\label{subsec:ap_forecast}

When creating our method for a two-component full-shape analysis, our first goal was to preserve the robust BAO information that we normally measure by re-scaling only the peak component. In order to check if our method succeeded in isolating this information, we fit the mock data using a BAO type model where we fix the smooth component, and only re-scale the peak. We then compare the posterior distributions of the BAO peak scale parameters ($\phi_\mathrm{p}$ and $\alpha_\mathrm{p}$) to the posteriors obtained from the full-shape analysis. The results are shown in the left panel of Figure \ref{fig:scale_params}. We show the constraints for a joint analysis of \lyalya\ and \lyaqso.  We find that our method arrives at BAO measurements in very good agreement with classic BAO analyses, which means that by re-scaling the smooth component we do not influence the measurement of the position of the acoustic peak. 

Our next goal for these forecasts is to understand the constraining power we have on the smooth component scale parameters, $\phi_\mathrm{s}$ and $\alpha_\mathrm{s}$. To this end, we consider a few different fitting strategies. As discussed above, before an actual measurement of these parameters, a full analysis of potential systematic errors needs to be performed. This study would inform the different analysis choices that need to be made in order to obtain robust measurements. One of these choices is the smallest scale that we fit. For past \lyaf\ BAO analyses, this has been chosen to be $r_\mathrm{min}=10\; h^{-1}$Mpc. This choice is not as important for BAO analyses because the BAO peak is a large-scale feature, and so, is not affected by small-scale contaminants. However, when attempting to measure scale parameters using the broadband component, these small scales have the potential to provide a lot of information. This is both because of the extra data points, and also because these data points at small separations have higher signal-to-noise. Therefore, we test a few different values of $r_\mathrm{min}$ that represent the range of possible options. We showcase the best case scenario where we are not affected by systematic errors all the way down to $10\;h^{-1}$Mpc, a worst case scenario where we have to cut the small scales and $r_\mathrm{min}=50\;h^{-1}$Mpc, and an intermediate case where we cut to $r_\mathrm{min}=30\;h^{-1}$Mpc. The lower value was chosen based on the value used by \lya\ BAO analyses, however, it might be too optimistic given current understanding of the \lya\ correlation functions (see Section \ref{sec:discussion} for discussion). On the other hand, the choice of the upper value was made because we might start to lose BAO information when removing scales above $50\;h^{-1}$Mpc \citep[][]{Kirkby:2013}.


The forecast broadband scale parameter results (again for \lyalya\ and \lyaqso) are shown in the right panel of Figure \ref{fig:scale_params} for different $r_\mathrm{min}$. Figure \ref{fig:scale_params} also highlights the difference in constraining power between the AP parameter and the isotropic scale parameter. It shows that, using the BAO peak, we can obtain very good measurements of the isotropic scale parameter, $\alpha_\mathrm{p}$ (the $68\%$ confidence region is at a precision of $\sim0.5\%$). However, we do not have very good constraining power when it comes to the AP parameter, $\phi_\mathrm{p}$, for which the $68\%$ confidence region is at a precision of $\sim1.6\%$. This is in contrast to the AP measurement from the smooth component, where even in the worst case scenario the $68\%$ confidence region is at a precision of $\sim0.9\%$, and in the best case scenario it is at $\sim0.3\%$. This shows the large potential gain in cosmological information from adding this AP measurement from the broadband. 

\subsection{Cosmological forecasts}
\label{subsec:ap_cosmo}

\begin{figure}
    \includegraphics[width=0.47\textwidth,keepaspectratio]{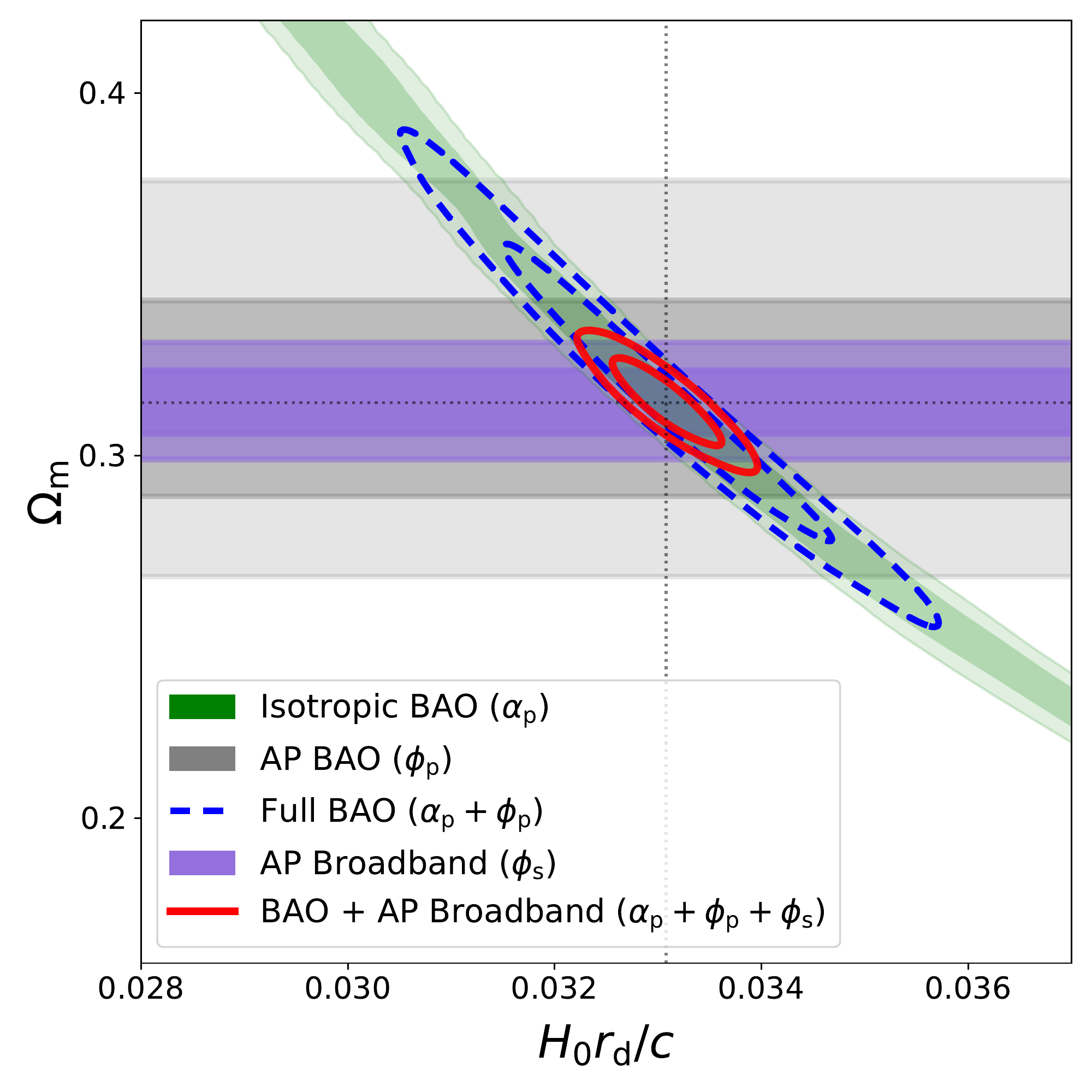}
    \caption{Forecast posterior distributions on cosmological parameters in flat \lcdm\ using different scale parameter measurements from the \lyaf\ correlation function. Measurements of the AP parameter ($\phi$) only constrain the matter fraction $\Omega_\mathrm{m}$, while the isotropic BAO scale measures $\Omega_\mathrm{m}$ and the combination $H_0 r_\mathrm{d}$. The AP measurement from the broadband ($\phi_\mathrm{s}$) is significantly better compared to the one from the BAO peak ($\phi_\mathrm{p}$). Therefore, the improved $\Omega_\mathrm{m}$ measurement leads to much tighter constraints when combined with the BAO measurement.}
    \label{fig:cosmo}
\end{figure}

We show the benefits of extracting more information from the \lyaf\ 3D correlation functions by performing a simple cosmological analysis using the forecast measurements obtained above. We use a flat \lcdm\ model, and we first model each of the measured parameters individually, in order to understand how each of them constrains cosmology. The cosmological interpretations of the scale parameters in terms of distances are given by Equations \ref{eq:phi} and \ref{eq:alpha}. Therefore, in order to complete our model, we just need the expressions for $D_\mathrm{M}$ and $H(z)$ in a flat \lcdm\ cosmology. The comoving angular diameter distance is given by:
\begin{linenomath*}
\begin{equation}
    D_\mathrm{M}(z) = c \int_0^z \frac{\mathrm{d}z'}{H(z')},
\end{equation}
\end{linenomath*}
and the Hubble parameter is given by the Friedmann equation:
\begin{linenomath*}
\begin{equation}
    \frac{H(z)^2}{H_0^2} = \Omega_\mathrm{m} (1 + z)^3 + \Omega_\Lambda + \Omega_\mathrm{r} (1 + z)^4.
\end{equation}
\end{linenomath*}
In flat \lcdm, the dark energy fraction can be computed from the matter and radiation fractional densities: $\Omega_\Lambda=1-\Omega_\mathrm{m}-\Omega_\mathrm{r}$. We also model the radiation fraction assuming a CMB temperature $T_\mathrm{CMB}=2.7255K$ \citep{Fixsen:1996,Fixsen:2009}, and a fixed neutrino sector \footnote{We use $N_\text{eff}=3.046$, with 2 massless species and one massive with $m_\nu=0.06$ eV that contributes to $\Omega_\mathrm{m}$.}. This means the only free parameters in $H(z)$ are $H_0$ and $\Omega_\mathrm{m}$.

For the AP parameter, we have a ratio of distances: $D_\mathrm{M}/D_\mathrm{H}$ (Equation \ref{eq:phi}), which means the Hubble constant cancels out. Therefore, in flat \lcdm, $\phi$ corresponds to a measurement of $\Omega_\mathrm{m}$. On the other hand, for $\alpha_\mathrm{p}$ we have a product of distances divided by the scale of the sounds horizon squared: $D_\mathrm{M} D_\mathrm{H} / r_\mathrm{d}^2$. As each of the two distances has a factor of $1/H_0$, we are left with the product $H_0^2 r_\mathrm{d}^2$, which means the two parameters are fully degenerate. Therefore, with $\alpha_\mathrm{p}$ we measure a combination of $\Omega_\mathrm{m}$ and the product $H_0 r_\mathrm{d}$. 

We use the $\alpha_\mathrm{p}$, $\phi_\mathrm{p}$ and $\phi_\mathrm{s}$ measurements presented above to constrain the relevant cosmological parameters. For $\phi_\mathrm{s}$ we use the result from the fit with $r_\mathrm{min}=30\;h^{-1}$Mpc, and we again use \texttt{PolyChord} to compute the posterior distributions. The constraints on $\alpha_\mathrm{p}$, $\phi_\mathrm{p}$ and $\phi_\mathrm{s}$ translate into measuring $H(z_\mathrm{eff})r_\mathrm{d}$ and $D_\mathrm{M}(z_\mathrm{eff})/r_\mathrm{d}$ with a precision of $\sim0.5-0.6\%$ each. In contrast, the DESI \lya\ BAO analysis is expected to measure $H(z_\mathrm{eff})r_\mathrm{d}$ and $D_\mathrm{M}(z_\mathrm{eff})/r_\mathrm{d}$ with a precision of $\sim0.9\%$\footnote{We recover this precision by translating the measurements of $\alpha_\mathrm{p}$ and $\phi_\mathrm{p}$ to $H(z_\mathrm{eff})r_\mathrm{d}$ and $D_\mathrm{M}(z_\mathrm{eff})/r_\mathrm{d}$, which validates our approach of re-scaling the covariance matrix presented in Section \ref{subsec:data}} \citep{Aghamousa:2016}. The cosmological parameter results using the individual measurements and their combinations are shown in Figure \ref{fig:cosmo}.

The constraint from the isotropic BAO measurement ($\alpha_\mathrm{p}$) leads to an elongated posterior with a strong degeneracy in the $\Omega_\mathrm{m} - H_0 r_\mathrm{d}$ space. This degeneracy is broken when combining with the $\Omega_\mathrm{m}$ constraint from $\phi_\mathrm{p}$ to obtain the usual anisotropic BAO measurement. However, as noted above, the AP measurement from the broadband is much better than the one measured from the peak. Therefore, by adding the $\phi_\mathrm{s}$ measurement to the BAO constraint, we can break the long correlation and obtain much better joint constraints. While the BAO measurements constrain $\Omega_\mathrm{m}$ and $H_0 r_\mathrm{d}$ with a precision of $8.9\%$ and $3.3\%$ ($68\%$ credible regions) respectively, adding the AP measurement from the broadband improves these constraints to $2.5\%$ and $1.0\%$.



\section{A joint analysis of the high-$z$ $3\times2$pt}
\label{sec:3x2pt}

In Sections \ref{sec:theory} and \ref{sec:lya_ap} we focused on extracting more information from the full shapes of \lyalya\ and \lyaqso\ through the AP parameter. We now turn our attention to the other source of cosmological information commonly used in full-shape analyses: redshift space distortions. In particular, we focus on the ability of joint analyses of the two \lya\ correlations to obtain meaningful measurements from RSD, and on the potential of a joint analysis of the three high redshift two point (high-$z$ $3\times2$pt) correlation functions: \lyalya, \lyaqso\ and \qsoqso.

\subsection{Context}

As the \lyaf\ velocity divergence bias, $b_{\eta,\mathrm{Ly}\alpha}$, is fully degenerate with the logarithmic growth rate, $f(z)$, we have so far treated RSD as a nuisance that we need to marginalize over. In practice, RSD analyses are sensitive to the combination $f\sigma_8$, where $\sigma_8$ is the amplitude of matter perturbations in spheres of 8 Mpc/h. This means that \lyalya\ effectively measures the combinations $b_{\mathrm{Ly}\alpha}\sigma_8$ and $b_{\eta,\mathrm{Ly}\alpha}f\sigma_8$.

The \lya-QSO cross-correlation could in theory be used to measure $f\sigma_8$. However, on its own it cannot constrain all the biases $(b_{\mathrm{Ly}\alpha},b_{\eta,\mathrm{Ly}\alpha},b_\mathrm{QSO})$ even for BAO analyses where we fix $f\sigma_8$ \citep[see Section \ref{subsec:cfmodel} and][]{Bourboux:2020}. On the other hand, a joint full-shape analysis of \lyaqso\ and \lyalya\ could help break these degeneracies and produce an $f\sigma_8$ constraint.

Another option for measuring $f\sigma_8$ at high redshift $(1.8<z<4)$ is to use the quasar auto-correlation, \qsoqso. The growth rate of structure was first measured from the quasar distribution by the eBOSS collaboration using the SDSS DR14 data \citep{GilMarin:2018,Zarrouk:2018,Hou:2018}. They performed full-shape analyses on both the 3D power spectrum and the 3D correlation function. With the last eBOSS analysis using SDSS DR16, these measurements have been updated and now provide a $\sim10\%$ constraint on the growth rate at an effective redshift $z_\mathrm{eff}=1.48$ \citep{Hou:2020,Neveux:2020}. The QSO sample contained $343,708$ quasars and spanned a redshift range of $0.8<z<2.2$. For comparison, DESI will measure about $1.7$ million QSOs at $z<2.1$ to be used as tracers only, and another $0.7$ million at $z>2.1$ to be used both as tracers and to measure the \lyaf\ \citep{Aghamousa:2016}.

The high redshift \qsoqso\ measurement could be combined with the two \lyaf\ correlations in a joint analysis. This could lead to improved $f\sigma_8$ constraints because of the information from the cross-correlation, and also due to the potential of the three correlations helping break parameter degeneracies. Therefore, our goal in this section is to study the potential of a high-redshift joint analysis of the three two-point (high-$z$ $3\times2$pt) correlation functions: \lyalya, \lyaqso\ and \qsoqso.

\subsection{Methods}
\label{subsec:3x2_methods}

We use a template linear power spectrum with a fixed normalization, which is proportional to $\sigma_8$. The logarithmic growth rate, $f(z)$, and $\sigma_8$ are completely degenerate in linear theory \citep{Percival:2009}, and therefore we are sensitive to the product $f(z)\sigma_8(z)$ for quasars and $b_{\eta,\mathrm{Ly}\alpha}f(z)\sigma_8(z)$ for the \lyaf. As $b_{\eta,\mathrm{Ly}\alpha}$ is unknown, we will continue sampling over the $\beta_{\mathrm{Ly}\alpha}$ parameter, effectively treating the \lyaf\ RSD term as a nuisance to be marginalized over.

We perform our analysis of the high-$z$ $3\times2$pt using the two-component full-shape method we introduced in Section \ref{sec:theory}. For the \lyaf\ auto and cross correlation we use the same simulated data and covariance matrices as described in Section \ref{sec:lya_ap}. For the QSO auto-correlation we also use a covariance matrix computed by \cite{Farr:2020}, re-scaled to the DESI area and number density as described in \ref{subsec:data}. The \qsoqso\ simulated correlation function is given by a fiducial model (no noise) following the best fit parameter values from \cite{Bourboux:2020}, again with the scale parameters set to unity.

Our effective parameter vector for joint fits is given by: $\{\phi_\mathrm{p},\alpha_\mathrm{p},\phi_\mathrm{s},\alpha_\mathrm{s},f\sigma_8,b_{\mathrm{Ly}\alpha}\sigma_8,b_\mathrm{QSO}\sigma_8,\beta_{\mathrm{Ly}\alpha},\sigma_\mathrm{v}\}$. When fitting individual correlations we follow the approach we took in Section \ref{sec:lya_ap}, of fixing the QSO bias and RSD terms for the cross-correlation.

\subsection{Breaking parameter degeneracies}
\label{subsec:fsig8}

\begin{figure}
    \centering
    \includegraphics[width=0.48\textwidth,keepaspectratio]{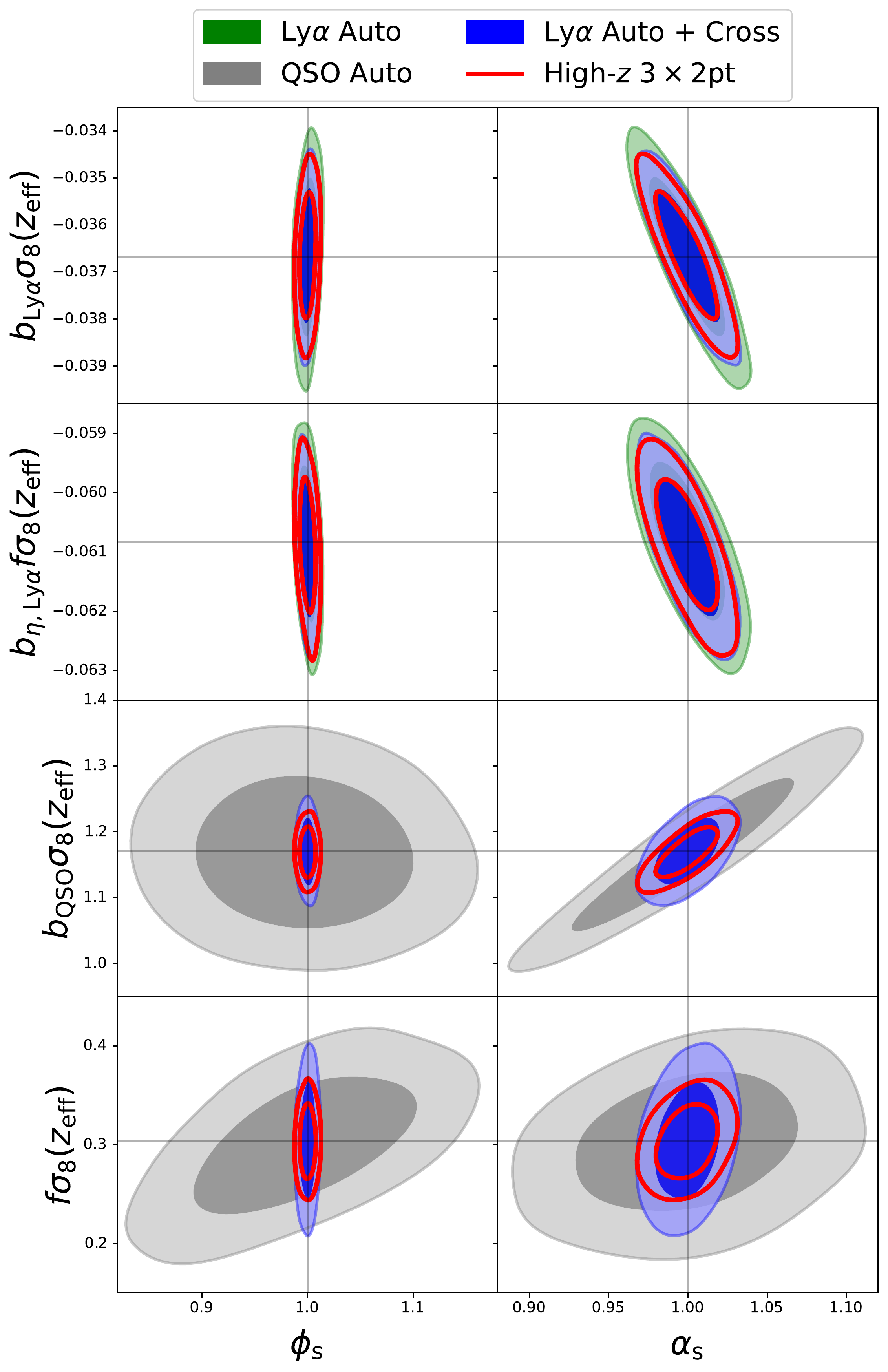}
    \caption{Posterior distributions of the \lyaf\ auto-correlation (green), the \lya\ auto + cross (blue), the QSO auto-correlation (gray), and the joint high-$z$ $3\times2$pt analysis of: \lyalya, \lyaqso\ and \qsoqso\ (red). We use a minimum separation $r_\mathrm{min}=30\; h^{-1}$Mpc. The first two rows show the parameters measured only by the \lyaf, while the bottom two rows show parameters constrained only by the quasar distribution. \lyalya\ and \lyaqso\ cannot constrain RSD individually, however, a joint full-shape analysis of both gives us an $f\mskip-4mu\sigma_8(z_\mathrm{eff}\simeq2.3)$ constraint that rivals the one from the quasar auto-correlation.}
    \label{fig:3x2pt}
\end{figure}

\begin{figure}
    \centering
    \includegraphics[width=0.5\textwidth,keepaspectratio]{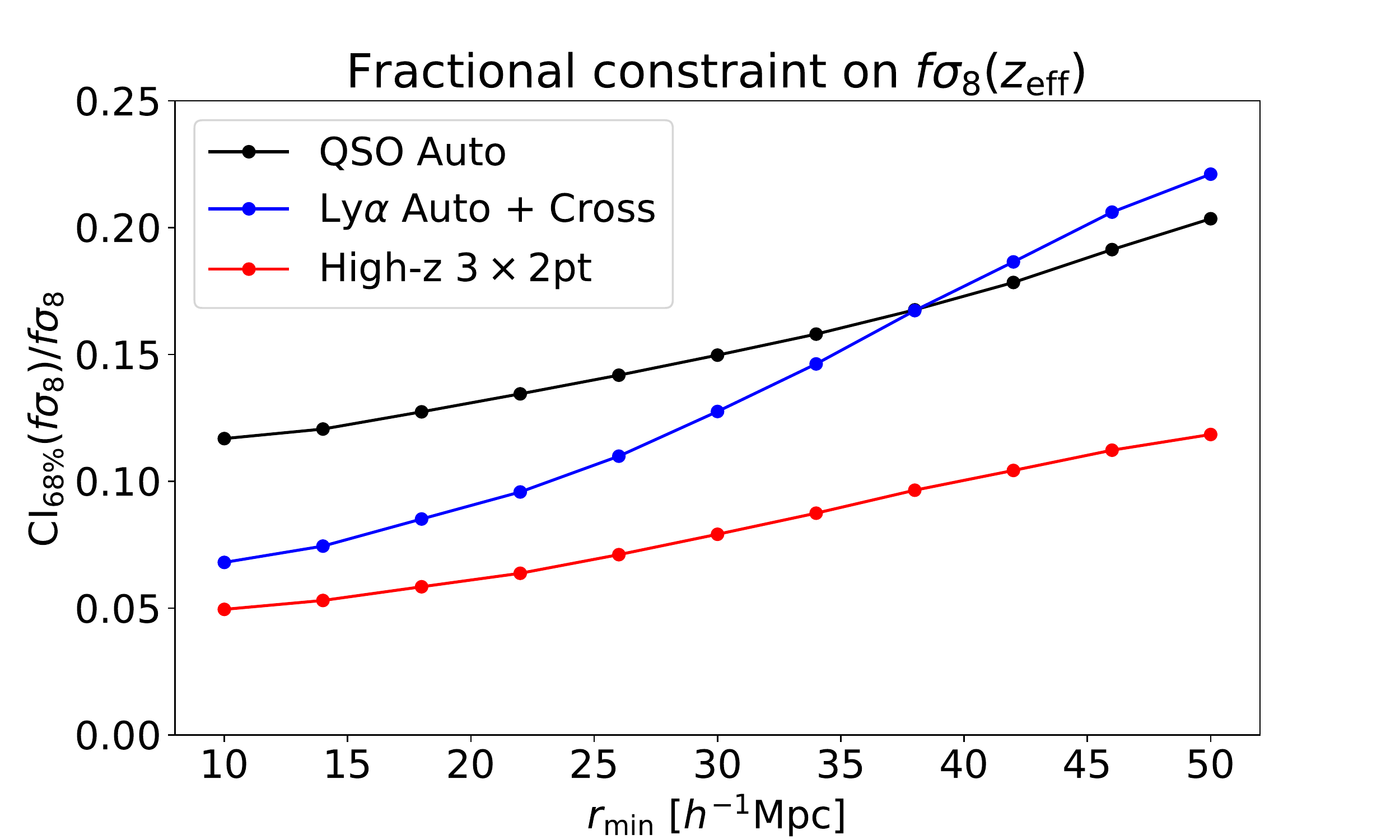}
        \caption{Forecast fractional constraints of the growth rate times the amplitude of fluctuations ($f\mskip-4mu\sigma_8$), as a function of the minimum separation ($r_\text{min}$) used for the fits. The black line shows the precision for the quasar auto-correlation, while the blue line shows the precision for a joint full-shape analysis of \lyalya\ and \lyaqso. The most precise and robust $f\mskip-4mu\sigma_8(z_\mathrm{eff}\simeq2.3)$ measurement is obtained by jointly fitting all three correlation functions (red line).
    }
    \label{fig:growth}
\end{figure}

In our parametrisation, the BAO parameters $(\phi_\mathrm{p},\alpha_\mathrm{p})$ are decoupled from the rest of the analysis. Therefore, as long as there is negligible cross-covariance between the different correlations, there is no benefit to $(\phi_\mathrm{p},\alpha_\mathrm{p})$ constraints from performing a joint analysis (i.e. fitting the correlations as one data vector). This has been the case so far with \lyalya\ and \lyaqso\ in BOSS and eBOSS \citep[e.g.][]{Bautista:2017,duMasdesBourboux:2017,Bourboux:2020}, but the cross-covariance for DESI remains to be studied.

The benefits of performing the joint analysis should be most pronounced when it comes to the parameters we measure from the full-shape analysis: $\phi_\mathrm{s}$, $\alpha_\mathrm{s}$ and $f\sigma_8$. This is firstly due to the fact that these parameters are correlated with some of the nuisance parameters, and therefore, a joint analysis would allow us to disentangle their effects and lead to improved constraints. This is illustrated in Figure \ref{fig:3x2pt} (where we use a minimum separation $r_\mathrm{min}=30 \; h^{-1}$Mpc). The top two rows show parameters that are only measured by the \lya\ auto-correlation, while the bottom two rows show parameters that are only measured by the QSO auto-correlation. Note that the fact that $\phi_\mathrm{s}$ does not seem to be correlated with $b_{\eta,\mathrm{Ly}\alpha}f\sigma_8$ in Figure \ref{fig:3x2pt} for \lyalya\ is just due to the scale of the axes which is set to display the weak \qsoqso\ constraint. AP and RSD are correlated, however, we do not expect these correlations to be the same for galaxies and the forest because the two tracers cluster differently (e.g. $\beta_{\mathrm{Ly}\alpha}\sim1.67$ while $\beta_\mathrm{QSO}\equiv f/b_\mathrm{QSO}\sim0.26$).

The cross-correlation requires all four parameters: $(b_{\mathrm{Ly}\alpha}\sigma_8,b_{\eta,\mathrm{Ly}\alpha}f\sigma_8,b_\mathrm{QSO}\sigma_8,f\sigma_8)$, however, the system is degenerate. On the other hand, when we run a joint analysis of the cross-correlation with the \lya\ auto-correlation (blue) we are able to constrain this system, because of the tight measurements of $b_{\mathrm{Ly}\alpha}\sigma_8$ and $b_{\eta,\mathrm{Ly}\alpha}f\sigma_8$ from \lyalya. This leads to a constraint on $f\sigma_8$ of $12.8\%$ ($68\%$ confidence region), which is tighter than the one from the QSO auto-correlation of $15.4\%$ (bottom left panel of Figure \ref{fig:3x2pt}).

The second benefit of performing this joint analysis is due to the correlation between RSD and the AP effect. When measuring $f\sigma_8$, we have to marginalize over the AP parameter. If we knew the true background cosmology, i.e. for fixed AP, we would obtain much better measurements of the growth rate. Even though the \lya\ auto cannot directly measure the growth rate, it constrains the AP parameters (especially $\phi_\mathrm{s}$) very precisely. Therefore, including \lyalya\ in a joint analysis with \qsoqso\ can help break the correlation between RSD and AP, and improve the $f\sigma_8$ constraint. This is illustrated in the bottom left panel of Figure \ref{fig:3x2pt}.

The joint high-$z$ $3\times2$pt analysis appears to work well in breaking parameter correlations when it comes to $f\sigma_8$ and $\phi_\mathrm{s}$. However, that is not the case with $\alpha_\mathrm{s}$. While performing a joint analysis does lead to better constraints on this parameter, the posterior remains very correlated with all three biases (right column of Figure \ref{fig:3x2pt}). This leaves a measurement of $\alpha_\mathrm{s}$ prone to systematic errors, and therefore supports our decision from Section \ref{sec:theory} not to focus on its cosmological interpretation.

So far in this section we used a minimum separation of $r_\text{min}=30 \;h^{-1}$Mpc to show how joint analyses help us break parameter degeneracies. However, we also want to test how these potential $f\sigma_8$ measurements would be affected if we could go to smaller scales (e.g. by having better models for non-linearities), or we had to cut even more data due to systematic effects on these scales. We show this in Figure \ref{fig:growth}, where we plot the marginalized fractional $68\%$ credible regions on $f\sigma_8$ from the QSO auto, the \lya\ auto $+$ cross, and the joint analysis of all three correlations. We find that when we can include data at small scales ($r_\text{min} \lesssim 35\;h^{-1}$Mpc), the \lya\ auto $+$ cross combination (blue line) gives us better constraints compared to the QSO auto (black line). On the other hand, if we have to cut the small scales ($r_\text{min} \gtrsim 35\;h^{-1}$Mpc), the \lya\ measurement degrades very fast, and the QSO auto becomes comparable and even slightly better at constraining $f\sigma_8$. This gives another advantage for performing a joint high-$z$ $3\times2$pt analysis, because it leads to much more stable and robust measurements (red line). While the quasar auto-correlation can constrain $f\sigma_8(z_\mathrm{eff}\simeq2.3)$ with a precision of $12-20\%$ depending on $r_\mathrm{min}$, the high-$z$ $3\times2$pt analysis can achieve a precision of $5-12\%$.

\begin{figure*}
    \includegraphics[width=1\textwidth,keepaspectratio]{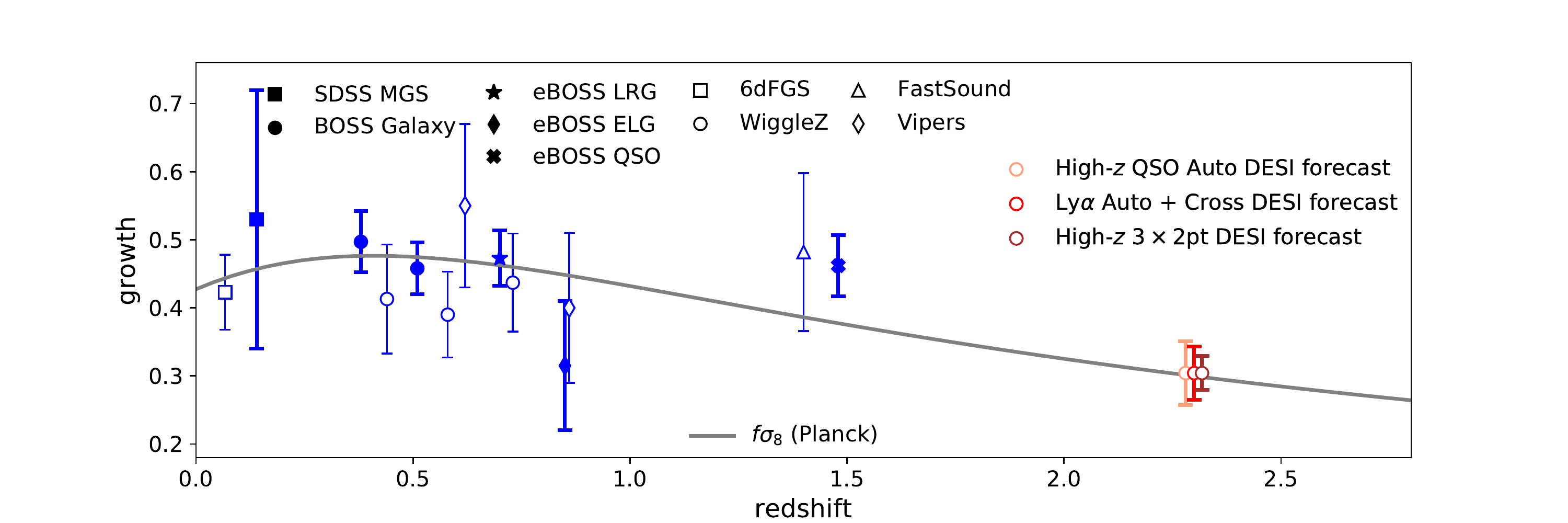}
    \caption{
    The growth rate times the amplitude of fluctuations ($f\mskip-4mu\sigma_8$) as a function of redshift. The gray line is the best-fit of CMB measurements from Planck. The blue points are some of the existing $f\mskip-4mu\sigma_8$ measurements. All of these measurements are at redshifts $z<2$, with most of them at $z<1$. The three points at high redshift are forecast constraints from DESI for the (high redshift) quasar auto-correlation, \lya\ auto and cross-correlations, and the joint high-$z$ $3\times2$pt analysis. Note that all three measurements are at the same effective redshift (given by the middle point), but are plotted at slightly different redshifts for visualization purposes.
    }
    \label{fig:gr_vs_z}
\end{figure*}

We have also checked how the two-component full-shape approach affects our results by comparing it with the approach usually taken in galaxy full-shape analyses of fitting the full correlation as one component (no peak/smooth decomposition). The $f\sigma_8$ constraints are larger when sampling four parameters (our two-component approach) versus two parameters (the one-component approach used in galaxy full-shape analyses). This is to be expected as the model has more degrees of freedom. However, the effect is very small when it comes to the high-$z$ $3\times2$pt constraints. We found that using a value of $r_\text{min}=30\;h^{-1}$Mpc, we obtain a precision of $7.9\%$ on $f\sigma_8$ with the two-component approach, while with the one-component approach we obtain a precision of $7.6\%$. This does not significantly affect our conclusions in this work, but the two-component approach might be more advantageous when the effects of contaminants are studied, as it decouples the peak from the broadband (see Section \ref{sec:discussion}).

Finally, in Figure \ref{fig:gr_vs_z} we emphasize how useful a full-shape high-$z$ $3\times2$pt analysis would be. We show in blue some of the current $f\sigma_8$ measurements from different surveys \citep{Ross:2015,Alam:2017,eBOSS:2020,Beutler:2011,Blake:2012,Okumura:2016,Pezzotta:2017}. All of these measurements are at redshifts $z<2$, with most of them at $z<1$. The three points on the right show our DESI forecasts of $f\sigma_8(z_\mathrm{eff})$ at an effective redshift $z_\mathrm{eff}\simeq2.3$. We use a conservative $r_\mathrm{min}=30 \; h^{-1}$Mpc. This analysis would allow us to study the growth rate of cosmic structures at higher redshifts than ever before.

The results in this section show the potential of a joint full-shape analysis of the three correlation functions: \lyalya, \lyaqso\ and \qsoqso, when it comes to measuring RSD and the AP effect. The next steps required for such an analysis are to improve the model by adding contaminants and better non-linear models, and to study the potential systematic errors that would affect this measurement, especially on the \lyaf\ side where a full-shape analysis of the 3D correlation function has never been done. We discuss these in more detail in the next section.

\section{Discussion and next steps}
\label{sec:discussion}

In this work we have shown the potential for extracting more cosmological information from the \lyaf\ 3D auto-correlation function and its cross-correlation with quasars. We took the template fitting approach where we use a template power spectrum to measure a few physically meaningful quantities that are easy to interpret and translate to cosmological constraints given some model. In our case, these quantities are the anisotropic scale parameter ($\phi$), the isotropic scale of the BAO peak ($\alpha_\mathrm{p}$) and the growth rate times the amplitude of fluctuations in spheres of $8\;h^{-1}$Mpc ($f\sigma_8$). This approach should simplify the study of the impact of contaminants because we only have to deal with a few parameters whose effect we understand very well.\footnote{This is in contrast to a direct fit of cosmological parameters where a study of contaminants would be much harder. This is due to the larger parameter space, but also because it is harder to identify and separate the effects of these parameters on the correlation function.} Such a study is required before a full-shape analysis of the \lyaf\ correlation functions is performed on real data, however, it is outside the scope of this work. Here we wish to briefly go over the most important contaminants, and mention what we can do to minimize their impact. In particular, the most relevant contaminants for the measurement of AP and RSD are those that introduce anisotropies.

High column density (HCD) systems are a significant contaminant for the \lyaf\ due to their broad absorption profile and long damping wings \citep{Font-Ribera:2012,Rogers:2018}. However, they also trace the underlying density field, which means they can add extra signal if modeled correctly. In past BOSS and eBOSS analyses, large damped \lya\ systems (DLA) that could be identified were masked \citep[e.g.][]{Bautista:2017,Bourboux:2020}. However, clustering measurements could potentially be biased by masking part of the spectrum as the mask is correlated to the density field. This was not a problem for BAO analyses, but its impact on a full-shape analysis needs to be tested. On the other hand, the small HCDs were left in the data and had to be included in the model. \cite{Rogers:2018} showed that HCDs can be successfully modeled down to the smallest scale considered in this work ($\sim10\;h^{-1}$Mpc), by using a simple model in linear theory, with a separate bias and RSD parameter, convolved with Voigt profiles for the damping wings.

The \lyaf\ auto and cross correlation functions are contaminated by metal transitions with rest-frame wavelength close to that of the \lya\ transition, that add correlations between themselves and the \lyaf\ or quasars. They are also contaminated by metal lines that are further away in rest-frame wavelength through their own auto-correlation. These metal lines have been successfully modeled for BOSS and eBOSS \citep{Bautista:2017,duMasdesBourboux:2017,DeSainteAgathe:2019,Blomqvist:2019,Bourboux:2020} by adding extra correlations with the same form as \lyalya\ and \lyaqso, and with their own bias and RSD parameters. However, these still need to be tested at DESI-level precision, and for the full-shape analysis we also have to test how the metal lines affect the measurement of the AP parameter and RSD.

Another important source of contamination are QSO redshift errors, which could introduce a systematic bias if not modeled correctly. Non-linear peculiar velocities also have big impact on the anisotropy because they create fingers of god. For \lyaf\ analyses (and in this work), these two effects have been modeled using simple damping terms with a Lorentzian or Gaussian profile based on \cite{Percival:2009}. For a full-shape analysis, we might need to use more complex models as was done for past quasar auto analyses \citep[e.g.][]{Hou:2020,Neveux:2020}. Quasar radiation effects (also known as the transverse proximity effect) are also an important source of contamination for the \lya-quasar cross-correlation. This is because the quasar radiation increases the ionization fraction in the surrounding gas, leading to less \lyaf\ absorption \citep{Font-Ribera:2013}. This effect has been modeled analytically and was shown to not have a significant impact on BAO analyses \citep{duMasdesBourboux:2017,Bourboux:2020}; however this needs to be tested for a full-shape analysis as well.

The final effect we consider is the fitting of the quasar continuum, which removes power on scales larger than the size of the forest. This produces a distortion in the measured correlations along the line of sight, and therefore introduces another source of anisotropy. This has been successfully modelled through a distortion matrix \citep{Bautista:2017} for BOSS and eBOSS. A similar approach could be sufficient for a full-shape analysis using DESI, but this needs to be tested.

All of the contaminants presented here have been studied before and are modeled in existing \lya\ BAO analyses. However, what still needs to be understood is how they interact with the new parameters we wish to study ($\phi_\mathrm{s}$ and $f\sigma_8$). Additionally, \cite{Bourboux:2020} found that adding broadband polynomials to the model can improve the fit of the correlations, which could point to contaminants that are not modeled well enough, or new effects that have not been considered. The addition of these polynomials was shown not to have a significant impact on BAO measurements, but they cannot be used for full-shape analyses because we want to extract broadband information, not marginalize over it. Therefore, a careful analysis on the impact of contaminants on AP and RSD measurements needs to be performed in order to determine if and on what scales current models are appropriate for a full-shape analysis of the \lyaf\ correlations. Furthermore, an analysis of potential systematic errors would inform the decisions related to which scale-parameters to sample (e.g. whether to have two separate $\phi$ parameters). We also mention that even in the worst-case scenario where we have to cut the small scales due to some significant systematic bias, we have shown that a full-shape analysis of the high-$z$ $3\times2$pt could still lead to state of the art cosmological measurements at redshifts $1.8<z<4$.

\section{Conclusions}

The Lyman-$\alpha$ (\lya) forest 3D auto-correlation function (\lyalya) and its cross-correlation with the quasar (QSO) distribution (\lyaqso) are currently some of the best cosmological probes of the Universe at redshifts $1.8<z<4$. However, so far they have only been used to measure the BAO scale. In this work we proposed to expand the cosmological information extracted from these statistics by fitting the full shape of these correlations in order to measure the Alcock-Paczynski (AP) parameter.

In Section \ref{sec:theory} we introduced our model for fitting the correlation function using a two-component approach, where we decomposed the template power spectrum into a peak component which contains the BAO information, and a smooth component. We then re-scaled the two components independently in order to decouple the measurement of the BAO peak from the rest of the analysis. In Section \ref{sec:lya_ap} we studied the potential for measuring the AP effect from the broadband of the \lyaf\ correlations. We used simulated correlation functions and mock DESI covariance matrices within a simple linear model with no contaminants. We showed that our two-component full-shape method successfully isolates the measurement of the BAO peak by comparing it to a BAO only analysis. Furthermore, we showed that using this idealized approach, a joint full-shape analysis of \lyalya\ and \lyaqso\ from DESI could measure the AP parameter at an effective redshift $z_\text{eff}\simeq2.3$ with a precision of $0.3\%-0.9\%$ ($68\%$ credible regions). Compared to the expected DESI \lya\ BAO constraint on AP, which is $\sim1.6\%$, we were able to obtain roughly two to four times better precision. In Section \ref{subsec:ap_cosmo} we showed how this measurement would help constrain cosmological parameters in a flat \lcdm\ model. In the conservative case where we fit to a smallest scale of $30 \; h^{-1}$Mpc, the inclusion of the AP measurement from the broadband gives us roughly three times better precision on the relevant cosmological parameters compared to the BAO measurement.

In Section \ref{sec:3x2pt}, we studied the potential for measuring the logarithmic growth rate times the amplitude of fluctuations in regions of $8\;h^{-1}$Mpc ($f\sigma_8$) at high redshift ($1.8<z<4$) using the DESI \lyaf\ and quasar position measurements. An $f\sigma_8$ measurement at redshifts $z\gtrsim1.6$ is unprecedented. Neither the \lya\ auto-correlation or the \lya-QSO cross-correlation can constrain $f\sigma_8$ independently, due to a degenerate system of parameters. However, in Section \ref{subsec:fsig8} we showed that their combination (\lyalya\ $+$ \lyaqso) is able to break these parameter degeneracies and obtain a measurement of $f\sigma_8(z_\mathrm{eff})$ at an effective redshift $z_\mathrm{eff}\simeq2.3$. This joint analysis was able to obtain constraints of $7\%-22\%$ ($68\%$ credible regions) depending on the minimum separation used. For comparison, with the high redshift quasar auto-correlation (\qsoqso) from DESI, we were able to obtain a precision of $12\%-20\%$. Furthemore, we showed that combining the two \lya\ correlations with the quasar auto-correlation in a joint analysis of the three high-redshift two-point correlation functions (high-$z$ $3\times2$pt) would give us the most precise and robust measurement of $f\sigma_8$ at these redshifts. We found that a high-$z$ $3\times2$pt analysis of the full DESI data could be able to measure $f\sigma_8(z_\text{eff}\simeq2.3)$ with a precision of $5\%-12\%$, depending on the minimum separation used.

In this work we have shown how to extract more information from the 3D distribution of the \lyaf\ through the AP parameter. We have also shown it is possible to measure $f\sigma_8$ through a joint full-shape analysis of \lyalya\ and \lyaqso. While the DESI \lya\ BAO analysis is expected to measure $H(z_\text{eff})r_\mathrm{d}$ and $D_\mathrm{M}(z_\text{eff})/r_\mathrm{d}$ with a precision of $\sim0.9\%$, adding the AP measurement from the broadband could give us constraints of $\sim0.5\%$. On the other hand, performing a high-$z$ $3\times2$pt analysis would allow us for the first time to measure $f\sigma_8$ at high redshift.

\section*{Acknowledgements}
AC was supported by a Science and Technology Facilities Council (STFC) studentship. AFR acknowledges support by FSE funds through the program Ramon y Cajal (RYC-2018-025210) of the Spanish Ministry of Science and Innovation. This work was partially enabled by funding from the UCL Cosmoparticle Initiative. SN acknowledges support from an STFC Ernest Rutherford Fellowship, grant reference ST/T005009/1.


\section*{Data Availability}
The data underlying this article will be shared on reasonable request to the corresponding author.



\bibliographystyle{mnras}
\bibliography{main} 








\bsp	
\label{lastpage}
\end{document}